\documentclass[aps,prl,twocolumn,floatfix,superscriptaddress,footinbib, sort&compress, numbers, merge, reprint]{revtex4-1}
\usepackage{graphicx}
\usepackage{amsmath}
\usepackage{amssymb}
\usepackage{booktabs}
\usepackage{color}
\usepackage{natbib}

\usepackage{chngcntr}

\usepackage[utf8]{inputenc}
\usepackage[tight,sf]{subfigure}%

\newcommand{\be}{\begin{eqnarray}}
\newcommand{\ee}{\end{eqnarray}}

\newcommand{\bs}{\mathbf}

\begin{document}

\title{Kibble-Zurek mechanism in curved elastic surface crystals}
\author{Norbert Stoop}
\affiliation{Department of Mathematics, Massachusetts Institute of Technology, 77 Massachusetts Avenue, Cambridge,~MA~02139-4307, USA}
\author{J\"orn Dunkel}
\affiliation{Department of Mathematics, Massachusetts Institute of Technology, 77 Massachusetts Avenue, Cambridge,~MA~02139-4307, USA}

\begin{abstract}
Topological defects shape the material and transport properties of physical systems. Examples range from vortex lines in quantum superfluids,  defect-mediated buckling of graphene,  and grain boundaries in ferromagnets and colloidal crystals, to domain structures formed in the early universe.  The Kibble-Zurek (KZ) mechanism describes the topological defect formation in continuous non-equilibrium phase transitions with a constant finite quench rate. Universal KZ scaling laws have been verified experimentally and numerically for second-order transitions in planar Euclidean geometries, but their validity for discontinuous first-order transitions in curved and topologically nontrivial systems still poses an open question. Here, we use recent experimentally confirmed theory to investigate topological defect formation in curved elastic surface crystals formed by stress-quenching a bilayer material. Studying both spherical and toroidal crystals, we find that the defect densities follow  KZ-type power laws independent of surface geometry and topology. Moreover, the nucleation sequences agree with recent experimental observations for spherical colloidal crystals. These results suggest that KZ scaling laws hold for a much broader class of dynamical phase transitions than previously thought, including nonthermal first-order transitions in non-planar geometries.
\end{abstract}

\maketitle

\textbf{Introduction.} 
Topological defects are localized perturbations that break the global symmetry of ordered solids or liquids. 
These defects influence the elastic, magnetic, electronic and optical properties in many natural and man-made systems. Examples range from the domain structures formed in the early universe~\cite{Kibble1976, Pospelov2013} to liquid crystals~\cite{Chuang1991, deGennes1995,Nikkhou:2015aa,Wang2016}, grain boundaries in ferromagnets and colloidal crystals~\cite{Weiss1906, Edwards2014},  and charge transport in graphene~\cite{2007Cortillo_EPL,yazyev2007defect} and superconductors~\cite{Zurek1985,Hendry94,dodgson1997vortices,PhysRevB.80.180501}. Topological defects can be created by varying a control parameter, such as temperature or a magnetic field, rapidly across a phase transition. Understanding the complex nonequilibrium dynamics induced by these nonadiabatic quenches remains an important theoretical challenge. Early progress in the characterization of topological defects was made by Kibble~\cite{Kibble1976} in 1976, while studying domain formation during the rapid cooling of the early universe~\cite{2016Hindmarsh}.  About a decade later, Zurek~\cite{Zurek1985} showed how the defect density is related to the quench rate in general second-order phase transitions; this important breakthrough also advanced significantly the understanding of topological defect formation in quantum superfluids and superconductors~\cite{1996Zurek_PhysRep}. Since then, the Kibble-Zurek (KZ) power-law scaling predictions were confirmed experimentally and utilized in a variety of systems, including liquid crystals~\cite{Nikkhou:2015aa,2017Fowler}, colloidal monolayers~\cite{Deutschlander2015}, ion crystals~\cite{Ulm2013}, Bose-Einstein condensates~\cite{Anquez2016}, superfluids~\cite{Hendry94,Ruutu96,PhysRevB.80.180501} and cold atomic clouds~\cite{Labeyrie2016}. However, since previous theory~\cite{Kibble1976,Zurek1985,Galla2003,PhysRevX.2.041022} and experiments focused primarily on continuous second-order transitions in planar Euclidean spaces, relatively little is known about the existence of KZ-type scaling laws for other classes of phase transitions and in more complex geometries. Here, we show that analogous  scaling laws hold for nonthermal discontinuous first-order phase transitions on simply and not-simply connected curved surfaces.

\begin{figure*}[t!]
	\begin{center}
		\includegraphics[width=0.95\textwidth]{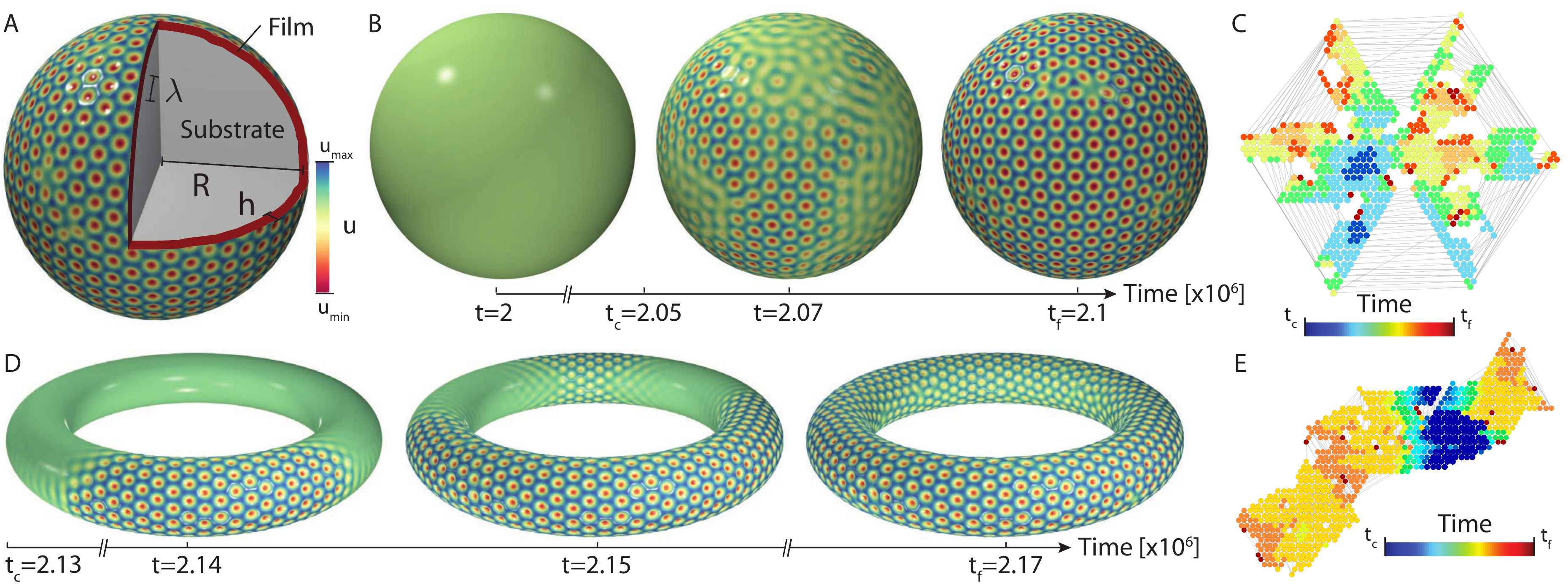}
	\end{center}
	\caption{Growth dynamics of elastic surface crystals in curved geometries. 
\textbf{A}:~Schematic of the simulated elastic bilayer material consisting of a thin film (thickness $h$) adhering to a soft substrate (radius $R$). Upon increasing the compressive film stress, the normal displacement $u$ of the film develops a hexagonal pattern with wavelength $\lambda$. 
\textbf{B}:~Surface crystallization dynamics on a sphere (radius $R/h=80$; Movie 1) for a slow quench $\mu=5\cdot 10^{-8}$. 
\textbf{C}:~Planar reconstruction of the crystallization process for the sphere in panel B. The growth of the elastic surface crystal from two initial nucleation sites (dark blue) proceeds along a predominantly regular hexagonal lattice structure, in close analogy with recent experimental observations for colloidal crystals on spherical liquid-liquid interfaces~\cite{meng2014elastic}.
\textbf{D}:~Crystallization dynamics on a torus  (radii $R/h=120$ and $r/h=24$; Movie 2)  for a slow quench ($\mu=5\cdot 10^{-8}$).
\textbf{E}:~The planar reconstruction for the torus in panel D reveals that toroidal surface crystals also grow sequentially along a regular hexagonal lattice structure, centered around wave-like geodesics of minimal absolute curvature. 
	}
	\label{fig1}
\end{figure*}

\par  
Topological defects in curved two-dimensional (2D) crystal structures arise in many biological and physical processes~\cite{Beller_PRE}, from plant growth~\cite{2013Newell_PRL,2013Saldoc,2015Newell_PhysD} and assembly of bacterial cell walls~\cite{Amir2014PNAS,AmirPRE2013}, viral capsids~\cite{Berger:1994aa,Lidmar2003PRE} and microtubules~\cite{Chretien:1992aa} to the targeted design of carbon nanotube sensors~\cite{2006Robinson_Nano} and microlense fabrication~\cite{Chan06Advanced}.  On closed manifolds, Euler's theorem~\cite{Euler} links the net charge of the topological defects to the genus $g$ of the underlying surface, imposing for example the 12 pentagons on a soccer ball. Depending on the details of the crystallization process (quench rate, geometric constraints, etc.), non-planar 2D crystals typically contain additional excess defects with zero total charge~\cite{giomi2008defective,Jimenez2016}. Recent studies provided important experimental and theoretical insights into the defect statistics and formation dynamics in 2D colloidal crystals assembled on curved liquid-liquid interfaces~\cite{vitelli2006crystallography, irvine2010pleats,meng2014elastic, kusumaatmaja2013defect}. Another promising class of experimental systems are curved elastic bilayer systems~\cite{Brojan14}, consisting of a soft substrate and a stiff surface film (Fig.~\ref{fig1}A), which can develop hexagonal wrinkling patterns under lateral compression induced by surface swelling~\cite{Breid13} or substrate depressurization~\cite{Denis14,Stoop15}.  Such elastic surface crystals allow the realization of nontrivial shapes of genus $g>0$, such as toroids, which are difficult to achieve in liquids. Moreover, because the transition from the unwrinkled to the hexagonal phase is of first order~\cite{Stoop15}, these soft matter systems also provide an ideal testbed to study generalizations 
of the KZ scaling laws.

\par 
The classical KZ argument for thermally induced second-order transitions builds on the fact that the correlation length $\xi$ and relaxation time $\tau$ diverge as $\xi\propto \vartheta^{-\nu}$ and   $\tau\propto \vartheta^{-z\nu}$ (critical slowing down), respectively, when the temperature $\vartheta$ is varied to drive the system continuously from the high-symmetry (e.g.~isotropic) phase to the lower-symmetry crystal phase. The critical exponents $\nu$ and $z$ encode the universality class of the transition. For a linear quench $\vartheta =\mu t$ with rate $\mu$, the system will not be able to relax defects during the time interval $|t|\lesssim t_f= \tau$, yielding the freeze-out condition $t_f \propto (\mu t_f)^{-z\nu}$ or, equivalently, $t_f\propto \mu^{-z\nu/(1+z\nu)}$.  The associated correlation length $\xi_f=\xi(t_f)\propto \mu^{-\nu/(1+z\nu)}$ implies the KZ scaling prediction for the defect density at freeze-out, $\rho_f\propto \xi_f^{-d}\propto \mu^{d\nu/(1+z\nu)}$, where $d$ is the space dimension. The analysis below shows that this argument can be generalized to stress-induced discontinuous pattern formation transitions observed on the 2D surfaces of curved elastic materials. Furthermore, our computations identify a dynamical analogy with recent experimental observations~\cite{meng2014elastic} on colloidal crystal formation in curved liquid-liquid interfaces.  Altogether, these results suggest that KZ-type scaling laws hold for a much broader class of dynamical phase transitions than previously thought. From a practical perspective, the subsequent analysis offers guidance for how to combine  quench dynamics and surface geometry to control both the frequency and localization of topological defects.

\section{Theory}

To investigate topological defect formation in curved geometries, we analyze an experimentally validated  continuum model~\cite{Stoop15} for the surface wrinkling in elastic bilayer materials consisting of a soft core and a stiffer outer shell (Fig.~\ref{fig1}A). This generalized Swift-Hohenberg (GSH) theory can reproduce quantitatively the experimentally measured equilibrium phase diagrams~\cite{Stoop15}, but its dynamical implications have not yet been explored.

\textbf{GSH theory for elastic surface crystals.} 
The GSH equations follow from the nonlinear Koiter shell theory~\cite{Ciarlet} by expanding the elastic energy of film and substrate in the dominant normal displacement field $u$~\cite{Stoop15}. Measuring length in units of the film thickness $h$, the surface energy functional of the GSH theory reads~\cite{Stoop15}
\begin{align} 
\mathcal{E} = \frac{k}{2} \int_{\omega} d\omega 
 \biggl[ \gamma_0 \left(\nabla u \right)^2  
+   \frac{1}{12} \left( \triangle u \right)^2 
+ a u^2 + \frac{c}{2} u^4  - \Gamma(u)  \biggr]
\notag
\end{align}
where $k=E_f/(1-\nu^2)$ for a film with Young's modulus $E_f$ and Poisson ratio $\nu$, and $d\omega$ is the surface element of the undeformed substrate. The nonlinear term 
$\Gamma(u)= \left[(1-\nu) b^{\alpha\beta}\nabla_{\alpha}u \nabla_{\beta}u + \nu b^\alpha_\alpha (\nabla u)^2 \right] u$ represents stretching forces to leading order in the curvature tensor $b^{\alpha\beta}$, with surface gradient $\nabla$ and Laplace-Beltrami operator $\triangle$ (throughout, Greek indices run over the set $\{1,2\}$ and Einstein's summation convention is used).
Taking the variation of $\mathcal{E}$ with respect to $u$ and assuming overdamped dynamics, the surface wrinkling process is described by the GSH equation~\cite{Stoop15, Jimenez2016}
\begin{equation}
	\tau_0 \frac{\partial}{\partial t} u = \gamma_0 \triangle u - \frac{1}{12} \triangle^2 u - a u -c u^3 - \delta_u \Gamma(u)
\label{e:GSHEgeneralized}
\end{equation}
where  $\gamma_0<0$, $c>0$ and $\delta_u \Gamma(u)$ denotes the functional derivative of the $\Gamma$-contribution to the energy functional~$\mathcal{E}$. Explicit expressions~\cite{Stoop15} for the coefficients and $\delta_u \Gamma(u)$ are summarized in the SI~\textit{Theory}. With no loss of generality, we measure time in units of the damping time~$\tau_0$ from now on, which is equivalent to setting $\tau_0=1$ in Eq.~\eqref{e:GSHEgeneralized}.
\par
Linear stability analysis of Eq.~\eqref{e:GSHEgeneralized} implies that wrinkling patterns form via a discontinuous transition (see Fig.~4 in Ref.~\cite{Stoop15}), when the control parameter $a$ falls below the critical value $a_c = 3 \gamma_0^2$, corresponding to an increase of the film stress $\sigma$ beyond the critical buckling stress $\sigma_c$. Defining the excess film stress $\Sigma_e = (\sigma/\sigma_c) -1$, the bifurcation parameter $a$ is related to $\Sigma_e$ by~\cite{Stoop15}
\begin{equation}
	a = a_c - \frac{3 c}{4} \Sigma_e\;.
\end{equation}
 In the regime beyond but still close to the wrinkling threshold, $0<\Sigma_e \ll 1$, nonlinear stability analysis confirms~\cite{Stoop15} that the wrinkling solutions adopt a hexagonal pattern (Fig.~\ref{fig1}) due to the $\delta_u\Gamma(u)$-term, which is breaking the $u\to-u$ symmetry of Eq.~\eqref{e:GSHEgeneralized}.

\begin{figure*}[t!]
	\begin{center}
		\includegraphics[width=0.95\textwidth]{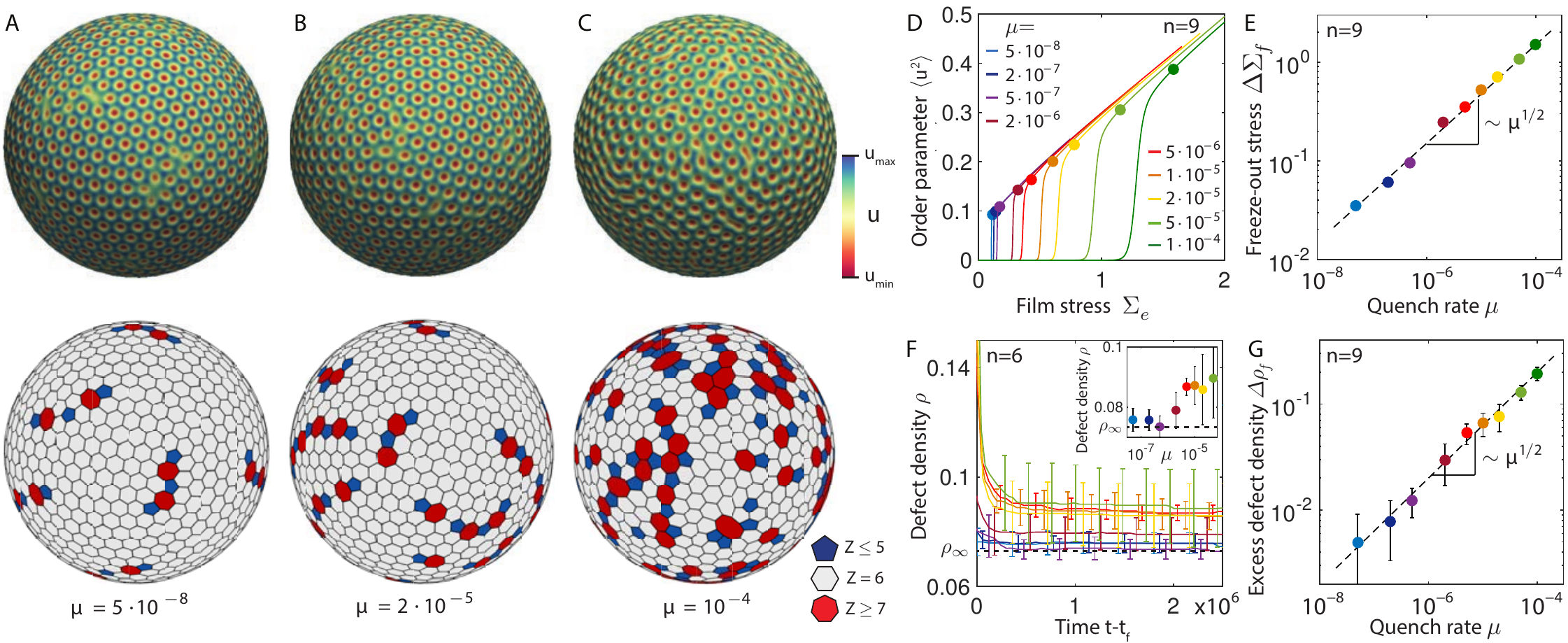}
	\end{center}
	\caption{KZ-type scaling laws for spherical surface crystals. 
\textbf{A}-\textbf{C}:~Crystalline surface patterns (top) and their corresponding Voronoi constructions (bottom) for different quench rates $\mu$ at freeze-out time $t_f$ show an increase in the defect density for fast quenches. 
\textbf{D}:~With increasing quench rate $\mu$, bifurcation of the order parameter $\langle u^2 \rangle$ becomes delayed, signaling non-adiabatic slowing down. Filled circles indicate the freeze-out film stress $\Sigma_{e,f}$ at which the system has resumed dynamics. 
\textbf{E}:~The net freeze-out stress $\Delta \Sigma_{f}=\Sigma_{e,f} - \Sigma_0$ follows a power-law scaling in the quench rate consistent with classical Kibble-Zurek predictions. Error bars are smaller than symbol size.
\textbf{F}:~When the quench is stopped at $\Sigma_{e,f}$, the system relaxes slowly to an equilibrium configuration by lowering its defect density, approaching the minimal equilibrium value $\rho_{\infty}$ in the adiabatic limit $\mu\to 0$ (inset).
\textbf{G}:~Although the pattern formation transition is of first-order, the excess defect density $\Delta \rho_f = \rho_f -\rho_{\infty}$ at freeze-out exhibits a square-root power law scaling.\label{fig2}
}
\end{figure*}

\par
\textbf{Stress-quenching of elastic surface crystals.}  
To obtain scaling predictions for the topological defect formation in elastic surface crystals, we first solve Eq.~\eqref{e:GSHEgeneralized} numerically  for linear stress quenches 
\begin{equation}\label{eq:quench}
	\Sigma_e(t) = \mu t
\end{equation}
with constant quench rate $\mu$,  driving the system from the unwrinkled to the hexagonal phase.  In all simulations, material parameters are chosen to match the experimental values reported in Refs.~\cite{Jimenez2016,Stoop15} (SI~\textit{Theory}). The three essential differences compared with classical KZ scenarios are that (i) the quench~\eqref{eq:quench} is nonthermal,  (ii) the wrinkling phase transition is of first order (Fig.~4b in Ref.~\cite{Stoop15}), and (iii) the underlying substrate geometries are non-planar.  
\par
To break the symmetry of the initially unwrinkled surface $u=0$, a small stationary random field $\epsilon$ with $|\epsilon| \ll 1$ is added to the rhs. of Eq.~\eqref{e:GSHEgeneralized} in simulations~(SI~\textit{Theory}). This $\epsilon$-inhomogeneity effectively models initial imperfections in the film displacement, thus mimicking realistic experimental conditions~\cite{Brojan14}.  We simulate Eq.~\eqref{e:GSHEgeneralized} for the linear quench \eqref{eq:quench} and a given realization of $\epsilon$ using the algorithm described in Ref.~\cite{Stoop15}. Numerical results presented below are averages over $n$ different realizations of~$\epsilon$, with $n$ specified on the corresponding graphs.

\section{Results}

To illustrate the effects of locally varying curvature and surface topology, we compare simulations for spherical and toroidal surfaces.

\textbf{Surface crystal growth under slow quenching.} 
For slow quasi-adiabatic quenches ($\mu\to 0$), we find that the crystallization process  is initiated at isolated nucleation sites and then spreads to cover the entire film (Fig.~\ref{fig1}B,D). This behavior is observed for both spheres and tori (Movies~1 and~2). Details of the spreading dynamics and local crystal orientation become evident by reconstructing the corresponding planar crystal patterns~(Methods). Starting from a random crystal site and one of its neighbors, we determine the relative positions of all other sites connected to this initial pair, resulting in the planar crystal representations of Fig.~\ref{fig1}C,E. The time evolution in these graphs shows how initially separated crystal patches merge to form a single connected crystal covering the entire surface. The radial cuts appearing in Fig.~\ref{fig1}C reflect the non-isometric character of the planar representation of spherical crystals,  whereas the toroidal crystals unfold nearly isometrically (Fig.~\ref{fig1}E).  One can see however  that, for both geometries, defects form only at the later stages when perfectly hexagonal crystal domains originating from different nucleation sites come in contact with each other. As discussed below,  qualitatively similar crystallization sequences were observed in a recent experiment~\cite{meng2014elastic} that studied the deposition of charged colloids onto a spherical oil-water interface. To connect with the ideas of Kibble~\cite{Kibble1976} and Zurek~\cite{Zurek1985}, we next use the GSH elasticity model to analyze the relation between topological defect formation and quench rate $\mu$, which is difficult to explore in curved colloidal systems due to experimental limitations on the particle deposition rates.

\textbf{KZ-type scaling in spherical surface crystals.}
We start our numerical scaling analysis by considering the case of globally constant curvature  as realized in spheres (Fig.~\ref{fig2}; Movie~1). 
In our simulations of Eq.~\eqref{e:GSHEgeneralized}, the quench rate is varied over the range $\mu \in [5\cdot10^{-8}, 10^{-4}]$,   consistent with the assumption of an overdamped dynamics.  Fixing the sphere radius $R/h=80$,  we characterize the transition from the unwrinkled to the wrinkled phase in terms of the average squared displacement $\langle u^2 \rangle$, where the brackets indicate an instantaneous surface average. In the adiabatic limit $\mu\searrow 0$, this order parameter vanishes  in the unwrinkled phase, $\langle u^2 \rangle =0$ for $\Sigma_e<0$, and jumps to a finite value $\langle u^2 \rangle >0$ when the excess film stress $\Sigma_e$ crosses zero from below. By contrast, for non-adiabatic quenches, we find that the system remains longer in the unwrinkled phase, before eventually breaking symmetry at some finite positive value $\Sigma_e>0$ (Fig.~\ref{fig2}D). Such delayed symmetry-breaking is also seen in the classical KZ mechanism, reflecting the critical slowing down in the relaxation dynamics near a thermal second-order phase transition. Yet, the wrinkling transition considered here is neither  thermal nor of second order, as hexagonal patterns arise through a subcritical bifurcation~\cite{Stoop15}.  To quantify the scaling behavior for this discontinuous first-order transition, we define the net freeze-out film stress $\Delta\Sigma_{f} =\Sigma_{e,f} - \Sigma_0$, with $\Sigma_{e,f}$ being the value at which the pattern amplitude is fully developed (Methods). The constant shift $\Sigma_0\approx 0.1$ is required to account for the material imperfections modeled by the $\epsilon$-inhomogeneity, as explained in Refs.~\cite{Erneux1986, Mandel1987}. Our simulations confirm power-law scaling $\Delta\Sigma_{f}\propto  \mu^{1/2}$  (Fig.~\ref{fig2}E), implying that the freeze-out time diverges as $t_f\propto\mu^{-1/2}$.

\par
The perhaps most interesting observable is the topological defect density $\rho_f$ at freeze-out $t_f$. Defects are crystal sites with coordination number $Z \ne 6$ and non-zero topological charge $s = Z - 6$. To identify the $\mu$-dependence of $\rho_f$ in the GSH theory, we determined the coordination number for each lattice site from the Voronoi cells of the displacement field $u$ (Methods). The resulting Voronoi graphs show that the freeze-out  density $\rho_f$ increases with the quench rate $\mu$ (Fig.~\ref{fig2}A-C). After a quench is completed, defect pairs are expected to annihilate by grain boundary movements. We tested this hypothesis in simulations by stopping the quench at the freeze-out value, $\Sigma(t)=\Sigma_{e,f}$ for $t>t_f$, so that the spherical surface crystals could relax to a stress equilibrium. For all considered quench rates, we find that the defect density approaches constant asymptotic values, which converge to the equilibrium value  $\rho_{\infty}$ as $\mu \searrow 0$ (Fig.~\ref{fig2}F). Note that, although the net toplogical charge is always $-12$ in agreement with Euler's theorem for hexagonal sphere tilings~\cite{Euler},  charge-neutral pairs of  penta- and heptagonal defects can reduce the elastic energy~\cite{bowick2002crystalline, bowick2000interacting, bausch2003grain, irvine2010pleats, Jimenez2016}, resulting in a non-zero defect density $\rho_{\infty}$ even at equilibrium. Defining the relative excess defect density at freeze-out $t_f$ by $\Delta\rho_f=\rho_f- \rho_{\infty}$, our numerical data is consistent with a KZ-type power law $\Delta\rho_f\propto \mu^{1/2}$ (Fig.~\ref{fig2}G).

\begin{figure*}[t!!]
	\begin{center}
		\includegraphics[width=0.95\textwidth]{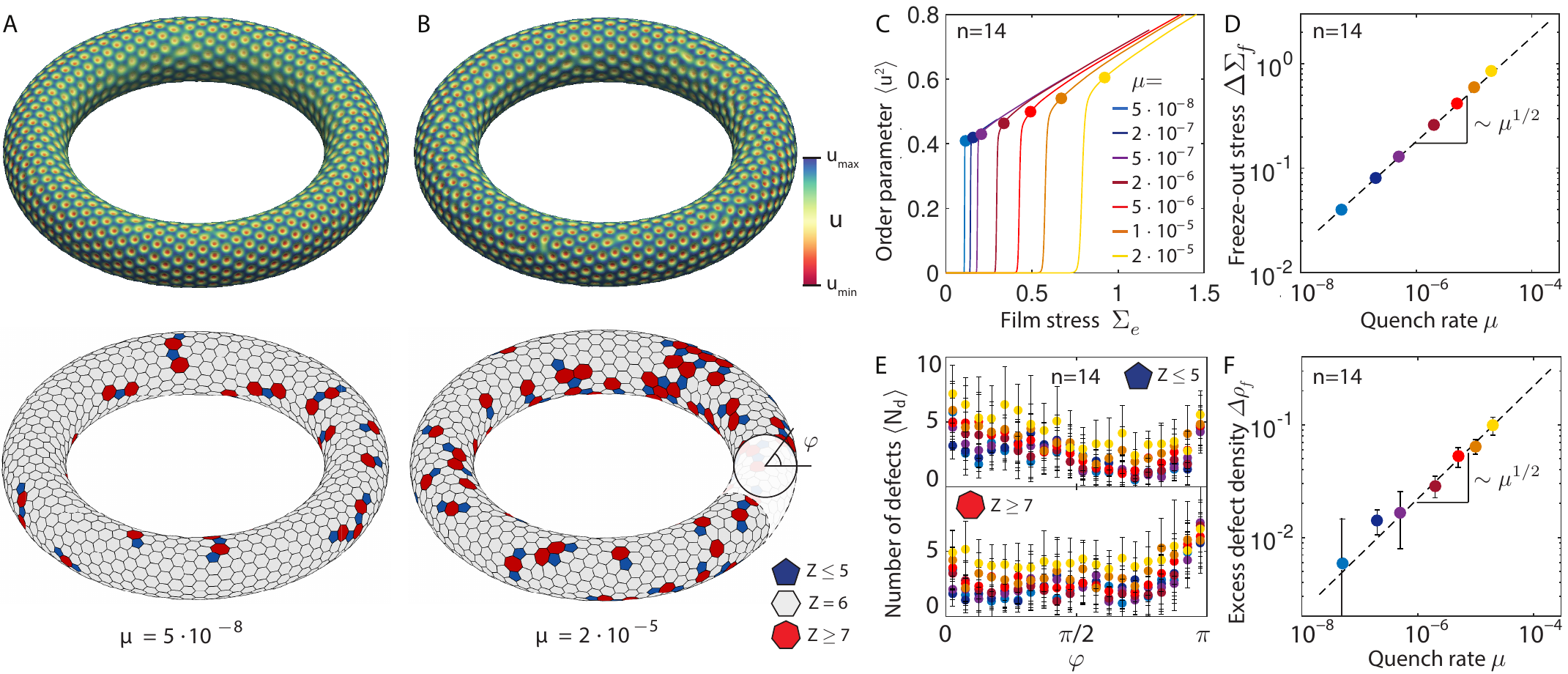}
	\end{center}
	\caption{KZ-type scaling laws for toroidal surface crystals. 
	\textbf{A},\textbf{B}:~Crystal structures (top) and Voronoi construction (bottom) for different quench rates $\mu$ at freeze-out time $t_f$ show an increase in the defect density for faster quenches. 
	\textbf{C}:~Delayed bifurcation of the order parameter $\langle u^2 \rangle$ for different quench rates, with the filled circles indicating the freeze-out film stress $\Sigma_{e,f}$. 
	\textbf{D}:~As for spherical substrates, we find $\Delta \Sigma_f \propto \mu^{1/2}$ for tori, confirming KZ-type scaling independent of surface genus and curvature. Error bars are smaller than symbol size.
	\textbf{E}:~The local average number of defects $\langle N_d \rangle$ at freeze-out time $t_f$ depends weakly on the angle $\phi$ along the minor radius ($\phi=0$ at the outer equator, and $\phi=\pi$ at the inner equator). 
	\textbf{F}:~The locally varying curvature does not affect the scaling law for the average freeze-out excess defect density $\Delta\rho_f=\rho_f-\rho_{\infty}$  which exhibits a square-root dependence on $\mu$.}
	\label{fig3}
\end{figure*}

\par
\textbf{KZ-type scaling for toroidal surface crystals.}
Local curvature variations can influence the equilibrium defect localization in curved elastic crystals~\cite{Jimenez2016}. To test if the defect scaling laws are also affected by curvature variations and surface genus, we performed additional parameter scans for tori. The locally varying curvature on a torus determines the strength of the symmetry-breaking term $\delta_u \Gamma(u)$ in Eq.~\eqref{e:GSHEgeneralized}, implying that a purely  hexagonal toroidal surface crystals exists only for sufficiently small aspect ratios $r/R$~\cite{Jimenez2016}.  We therefore focus on thin tori with $r/R=0.2$ (Fig.~\ref{fig3}A,B; Movie~2). Adopting the same small random $\epsilon$-inhomogeneities as for the sphere simulations (and, hence, the same shift value $\Sigma_0$), we observe that the freeze-out stress $\Delta\Sigma_f \propto \mu^{1/2}$  and  the freeze-out time $t_f\propto\mu^{-1/2}$ scale exactly as in the case of constant curvature (Fig.~\ref{fig3}C,D).  Furthermore, one again finds  that faster quenches lead to a higher density of defects at freeze-out (Fig.~\ref{fig3}A,B,F). Although Euler's theorem~\cite{Euler} imposes a vanishing total defect charge $S=0$ on a torus, excess defects of opposite charge tend to aggregate at the inner ($Z=7$) and outer ($Z=5$)  equators to lower the elastic energy of the toroidal crystal~\cite{Jimenez2016, bowick2004curvature, giomi2008elastic,giomi2008defective}, resulting in a non-zero equilibrium defect density $\rho_\infty$.  As in the sphere case, we find that the relative  excess defect density $\Delta\rho_f=\rho_f- \rho_{\infty}$ of the  toroidal surface crystals follows a square root law $\Delta\rho_f \propto \mu^{1/2}$ (Fig.~\ref{fig3}F), suggesting that local curvature variations and surface genus do not significantly affect the scaling laws. In the next part, we will rationalize these observations by considering the structure of the amplitude equations~\cite{Doelman2003Propagation} for the underlying GSH theory.

\section{Discussion}

\textbf{KZ-type scaling for nonthermal quenches.}
As outlined in the introduction, the original KZ scaling relations for continuous second-order transitions can be obtained by analyzing  how correlation length and relaxation time diverge as a function of the control parameter as one approaches the critical point from the high-symmetry phase.  By contrast, for the nonthermal first-order wrinkling transitions considered here, the correlation length is not defined in the unwrinkled phase, so that the standard KZ arguments do not apply. One can nevertheless explain the numerically determined  scaling laws by inspecting  the amplitude equations for the GSH model. To this end, we assume approximate hexagonal solutions  of the form $u=U(t)\sum_{a=1}^3 [e^{i\bs k_a\cdot \bs x}+ e^{-i\bs k_a\cdot \bs x}]$, where $\bs k_1=k_c (1,0)$, $\bs k_2=k_c(-1/2,  \sqrt{3}/2)$, $\bs k_3=k_c(-1/2, -\sqrt{3}/2)$ and $k_c = \sqrt{6|\gamma_0|}$. Inserting this ansatz into the GSH equation~\eqref{e:GSHEgeneralized}, one finds to leading order in $U$ and curvature (SI~\textit{Amplitude Equations}) 
\be
\frac{d}{dt} U\approx \frac{c \mu t}{12 \gamma_0^4}  U.
\ee
Dividing by $\sqrt{\mu}$ and defining a rescaled time $t'=\mu^{1/2} t$, we can remove the quench rate dependence at leading order, to obtain 
\be
\frac{d}{dt'} U\approx \frac{c t'}{12 \gamma_0^4}U.
\label{eq:muinvariance}
\ee
Non-autonomous equations of this type have been extensively studied in dynamical systems theory and are known to describe delayed bifurcations with some characteristic delay time scale $t'_f$~\cite{Kuehn2011, Erneux1986, Baer1989,Mandel1987}. Identifying this characteristic delay with the freeze-out time implies $t_f\propto \mu^{-1/2}$, in agreement with our numerical results. It may be worth emphasizing again that this power-law scaling is a direct consequence of the nonthermal stress-quench dynamics. This example corroborates  that KZ-type scaling can occur even when the conventional critical slowing down of thermal systems is replaced by other delayed bifurcation mechanisms~\cite{2017Vella_NPhys}. Moreover, since the quench enters through a purely local $u$-term in Eq.~\eqref{e:GSHEgeneralized}, the scaling law is not affected by curvature variations. Finally, in order to explain the  scaling law for the excess defect density $\Delta \rho_f$ at freeze-out, it suffices to assume that the mean-squared distance $\delta^2$ between defect pairs grows diffusively,  consistent with quasi-random migration of defect precursors, so that at freeze-out  $\delta^2\propto  t_f\propto \mu^{-1/2}$. For a 2D surface, this then implies that $\rho_f\propto \delta^{-2} \propto \mu^{1/2}$,  in agreement with the numerical results in Fig.~\ref{fig2}G and~\ref{fig3}F.

\textbf{Universal nucleation dynamics in curved surface crystals.} 
Crystal growth in planar Euclidean geometry is well understood. By contrast, the complex interplay between kinetics, substrate curvature and defect formation is still being investigated~\cite{meng2014elastic,Beller_PRE}. The current interest in these topics is in parts driven by recent technological advances in the fabrication of graphene~\cite{chuvilin2010direct} and carbon nanotubes~\cite{2006Robinson_Nano} and by the development of modern confocal imaging techniques that make it possible to track micron-sized colloids and cells at high spatio-temporal resolution. For instance, the formation of hexagonal crystal structures similar to those described above can be observed during the early developmental stages in the fruit fly \textit{Drosophila melanogaster}, when nuclei migrate to accumulate underneath  the surface of the ellipsoidal shell that encapsulates the embryo~\cite{Tomer2012}. Similarly, ciliated somatic cells form a spherical crystal on the surface of the colonial alga \textit{Volvox carteri}~\cite{Drescher2010}, with the cells' arrangement determining the phototactic properties of the organism. 
\par
Important insights into the kinetics of crystal growth on curved substrates were obtained recently in a joint experimental and theoretical study~\cite{meng2014elastic} on the assembly of charged colloids on spherical liquid-liquid interfaces. The crystal formation dynamics observed in these experiments shares striking similarities with the nucleation sequences of the hexagonal wrinkling patterns shown in Fig.~\ref{fig1}C. In both cases, one first observes the formation of several smaller highly regular crystal patches, while the defects form during the later stages when two or more of these patches merge. These kinetic parallels suggest a certain universality in the crystal formation processes on curved surfaces in the slow-quench regime. Extrapolating these similarities to higher quench rates, one may hope that the scaling results identified here translate to other physical and biological systems that develop crystalline structures on their curved surfaces.

\section{Conclusions}

The above analysis shows that the KZ mechanism for continuous second-order transitions has a direct analog in first-order phase transitions, if the underlying amplitude equations exhibit delayed bifurcations. As a specific example, we have identified the power law scaling relation between topological defects densities and linear quench rates  for an experimentally validated generalized Swift-Hohenberg theory~\cite{Stoop15} describing stress-induced discontinuous surface wrinkling transitions in thin stiff films adhered to a curved soft substrates. With regard to applications, these scaling relations offer concrete guidance for controlling the number of topological defects by tuning the quench rate, extending previous work~\cite{giomi2008defective,Jimenez2016} that showed how substrate geometry can be used for defect localization.  The nucleation sequences leading to crystalline hexagonal surface patterns in the thin-film model are qualitatively similar to those reported recently for colloidal crystals on spherical liquid-liquid interfaces~\cite{meng2014elastic}. This suggests that elastic surface crystals, which have been realized by substrate depressurization~\cite{Brojan14} or surface swelling~\cite{Breid13}, can offer a flexible testbed for exploring generic aspects of crystal growth kinetics and topological defect formation in complex geometries.

\subsection*{Methods}

\textbf{Pattern analysis.} 
Simulations were performed using the algorithm detailed in Ref.~\cite{Stoop15}, adopting parameters for centimetre-sized polydimethylsiloxane-coated elastomer hemispheres~\cite{Denis14,Stoop15}. To reconstruct the curved crystal structure and detect topological defects, we first threshold the amplitude field $u$ obtained from simulations to find the center of each crystalline lattice site. Each site is then connected to its nearest neighbors via a Delaunay triangulation, and the hexagonal crystal structure as well as defects are obtained from the dual Voronoi graph. To find the flat crystal representation of Fig.~\ref{fig1}C,E, we first construct the Voronoi cells of the crystal. We then randomly select a regular crystal site $s_0$ and its six neighbors as center of the planar lattice. Based on their  positions relative to $s_0$,  the neighbors can be assigned to one of the six surrounding unit lattice positions. We then repeat this construction sequentially for each neighbor by aligning their closest neighbors with the existing planar lattice structure, following the procedure used in Ref.~\cite{meng2014elastic}. To obtain the time evolution, we first construct the final planar crystal from to the fully crystallized configuration at freeze-out time $t_f$. For all earlier times $t<t_f$, we use a Hungarian matching algorithm~\cite{Munkres} to track identical Voronoi sites and identify them in the fully crystallized planar configuration.

\textbf{Determination of freeze-out stress.} 
We first determine the stress $\Sigma_{\textrm{max}}$ and corresponding time $t_{\textrm{max}}=\Sigma_{\textrm{max}}/\mu$ where $\langle u^2 \rangle$ has largest slope. 
Although \mbox{$\Sigma_{\textrm{max}}-\Sigma_0 \propto \mu^{1/2}$} (Fig.~S1A,B), crystalline patterns are not yet fully developed at this value of the film stress, rendering an analysis of the defect density scaling practically unfeasible. To obtain robust estimates for the defect densities, we therefore add a time delay to $t_{\textrm{max}}$  to allow the hexagonal patterns to develop completely. According to Eq.~\eqref{eq:muinvariance}, the dynamics is rate-independent under the rescaling $t' = \mu^{1/2} t$ to first order (Fig.~S1C). Therefore, in order to give each quench the same relative amount of time for developing hexagons, we choose a time delay proportional $\mu^{-1/2}$, $t_f = t_{\textrm{max}} + \mu^{-1/2}\Delta\tau $, with $\Delta\tau=30$ kept constant throughout our analysis (SI~\textit{Freeze-out}). Accordingly, the freeze-out film stress is defined as $\Sigma_{e,f} = \mu t_f =  \Sigma_{\textrm{max}} + \mu^{1/2} \Delta\tau$.


\end{document}